# Edge Conditions for the Junction of Two Resistive Half-Planes with Different Surface Impedances

Igor M. Braver[a], Pinchos Sh. Fridberg[b], Khona L. Garb[c], Iosif M. Yakover[c]

[a]Sholom Aleichem ORT High School, Justiniskiu str. 65, Vilnius, LT-05100, Lithuania
[b]Retired professor of applied electrodynamics, Smelio str. 4, Vilnius, LT-10324, Lithuania
[c]Tel Aviv University, School of Electrical Engineering, Ramat Aviv, Tel Aviv, 69978, Israel
Corresponding author Garb Khona, khona@post.tau.ac.il

*Abstract*—**This work presents an analysis of the behavior of an electromagnetic field near the common edge of two resistive half-planes with different surface impedances. Contrary to the case of a single resistive half-plane, in the case of the impedance junction, *both* electric and magnetic fields' transverse components simultaneously contain a logarithmic singularity. It is shown that a surface current density has a finite jump that is proportional to the difference between the inverse impedances.**

## I. Introduction

In formulating the boundary value problems of electrodynamics describing structures with sharp edges, the uniqueness of the solution can be established only if the so-called edge conditions [1, 2] are satisfied. This condition requires that the electromagnetic energy stored in any finite neighborhood of the edge must be finite. In other words,

$$\lim_{V \to 0} \int_V dv \left( \varepsilon |\mathbf{E}|^2 + \mu |\mathbf{H}|^2 \right) = 0, \qquad (1)$$

where $\varepsilon$ and $\mu$ are the permittivity and permeability of the medium and $\mathbf{E}$, $\mathbf{H}$ are the vectors of electric and magnetic field respectively. It follows from (1) that all of the electromagnetic field components should grow more slowly than $1/\rho$, as $\rho \to 0$ ($\rho$ is the distance from the wedge). More detailed knowledge of the behavior of $\mathbf{E}$, $\mathbf{H}$ is not necessary for formulating of the problem. However, in numerical calculations the convergence of the solution may be improved if a known field behavior is included in the formulation of the problem, e.g., the well-known singularity near a metallic edge. There may also be high-power situations, where knowledge of field strength is needed in order to avoid breakdown.

Meixner [1, 2] found a series expansion of all components of electric and magnetic fields. Later authors, Mittra and Lee [3], Hurd [4], Van Bladel [5], who treated various different configurations, have followed Meixner. Meixner's method leads formally to a solution in the form of series containing powers of the distance from the edge. A recurrent system of ordinary differential equations arises for the coefficients of expansion, and so solving it we obtain a representation of the field in the vicinity of the edge of the structure. However, in certain situations the problem of finding the coefficients of the series becomes insoluble [6], which led the authors of [6] to the conclusion that Meixner's theory has an internal inconsistency. Makarov and Osipov [7] while investigating the dielectric structures have proved that the series describing the field near the edge of a dielectric wedge should contain the logarithmic functions in all cases where among the solutions of the characteristic equation there are roots that differ by integer $l$. In our papers [8–10] investigating the behavior of the electromagnetic field near the edge of a resistive half-plane, taken separately, as well as in conjunction with a perfectly conducting half-plane, it was proved, that using only series of $\rho$ in the expansion of the longitudinal to the wedge components of electromagnetic field leads simply to the trivial solution and new terms, containing logarithmic functions, should be added. We came to the same conclusions in [11, 12] by the study of the behavior of electromagnetic field near the common edge of a perfectly conducting wedge and a resistive half-plane. In paper [8] we were the first who investigated the structure for which the nature of electromagnetic field's singularity is not algebraic, but *logarithmic*. Such a structure comprised a resistive half-plane, either taken separately [8], or in conjunction with a perfectly conducting half-plane [9]. Afterward logarithmic singularity was mentioned (without investigating of fields' distribution) for the modified conducting half-plane and for the magnetically conducting half-plane [13, 14]. In all configurations above logarithmic singularity arises either only for electric $\mathbf{E}_\perp$, or only for magnetic $\mathbf{H}_\perp$ transversal with respect to the edge electromagnetic field's component. The present communication is aimed to investigate the junction of two resistive half-planes with different surface impedances. It will be shown that for such a structure both $\mathbf{E}_\perp$ and $\mathbf{H}_\perp$ contain the logarithmic singularity simultaneously. The present paper is organized as follows. In Section II we describe the statement of the problem, in Section III we investigate the behavior of electromagnetic field near the junction. Finally, Conclusions are given in Section IV.

## II. Statement of the Problem

Let us consider a resistive plane $y = 0$ (Fig. 1) with inhomogeneous surface impedance $W$. We assume that $W = W_1$ for $x > 0$ and $W = W_2$ when $x < 0$. On the surface of the resistive plane two-sided impedance type boundary conditions are fulfilled:

$$E_x^+ = E_x^- = W(H_z^+ - H_z^-),$$

$$E_z^+ = E_z^- = -W(H_x^+ - H_x^-),\quad (2)$$

where $W$ is surface impedance, and "+" and "–" denotes a field at $y = +0$ and $y = -0$ respectively. Now we consider two cases, having different symmetries of tangential magnetic field about the plane $y = 0$. In the even case, we obtain the trivial solution when the presence of resistive plane does not distort the electromagnetic field, because tangential components of **E** on its surface vanish. In the odd case ($H_x$, $H_z$, $E_y$ are odd functions and $E_x$, $E_z$, $H_y$ are even functions with respect to $y$) the boundary conditions can be written as

$$E_x = 2WH_z,\quad E_z = -2WH_x.\quad (3)$$

After that it will suffice to investigate the field only in the upper half-space $y \geq 0$.

While studying the behavior of the field at $\rho \to 0$ we can disregard the dependence of **E** and **H** on $z$ coordinate and consider two independent problems of defining $E_z$, $H_x$, $H_y$ (TM polarization) and $H_z$, $E_x$, $E_y$ (TE polarization). Dependence of the fields on time $t$ is chosen in the form $\exp(-i\omega t)$. We will find the distribution of the fields **E**, **H** satisfying (in the source-free region) Maxwell's homogeneous equations and boundary conditions (3). In both TM and TE polarizations, the components to be found are expressed as their scalar function $\chi$ ($\chi = E_z$ or $\chi = H_z$) which is the solution of the two-dimensional Helmholtz equation

$$\frac{1}{r}\frac{\partial}{\partial r}\left(r\frac{\partial \chi}{\partial r}\right) + \frac{1}{r^2}\frac{\partial^2 \chi}{\partial \varphi^2} + \chi = 0,\quad (4)$$

where $r = k\rho$, $k$ is the wavenumber in free space and is assumed that $k \neq 0$.

According to the method of papers [8, 9] function $\chi$ should be sought for in the double series form

$$\chi = r^\tau \sum_{m=0}^{\infty}\sum_{n=0}^{m} C_{mn}(\varphi) r^m \ln^n r,\quad (5)$$

where $\tau \geq 0$ — parameter to be found, $C_{mn}(\varphi)$ unknown functions on $\varphi$.

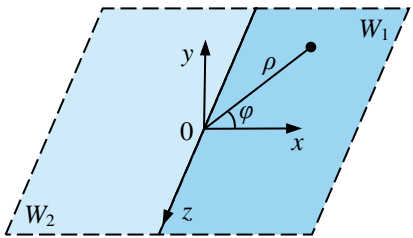

Fig. 1. Junction of two resistive half-planes.

The requirement $\tau \geq 0$ assures quadratic integrability (over the area with surface element $\rho d\rho d\varphi$) of the transverse field components in the vicinity of the junction. From this point of view $\tau > 0$ are acceptable at any $C_{mn}$ and $\tau = 0$ only if $\partial C_{mn}/\partial \varphi = 0$. Substitution of (5) in (4) leads to an infinite set of differential equations:

$$\partial^2 C_{mn}/\partial \varphi^2 + (m+\tau)^2 C_{mn} + 2(m+\tau)(n+1)C_{m,n+1}$$
$$+(n+1)(n+2)C_{m,n+2} + C_{m-2,n} = 0.\quad (6)$$

From here on $C_{mn} = 0$, if $m < 0$ or $m < n$. Constants of integration arising in solving (6) and the value $\tau$ which ensures nonzero solution are to be determined from boundary conditions (3). Substituting (5) also into (3) gives the equalities

$$\partial C_{mn}/\partial \varphi|_{\varphi=0} = -i(\varsigma/2W_1)C_{m-1,n}|_{\varphi=0},$$
$$\partial C_{mn}/\partial \varphi|_{\varphi=\pi} = i(\varsigma/2W_2)C_{m-1,n}|_{\varphi=\pi},\text{ if } \chi = E_z\quad (7a)$$

$$\partial C_{mn}/\partial \varphi|_{\varphi=0} = -i(2W_1/\varsigma)C_{m-1,n}|_{\varphi=0},$$
$$\partial C_{mn}/\partial \varphi|_{\varphi=\pi} = i(2W_2/\varsigma)C_{m-1,n}|_{\varphi=\pi},\text{ if } \chi = H_z,\quad (7b)$$

$\varsigma$ — is the wave impedance in free space. It is easy to see, that equations (7) and therefore final expansion of function $\chi$ for TM and TE polarizations transforms one from the other, when $\varsigma/2W$ is replaced by $2W/\varsigma$.

### III. FIELD INVESTIGATION

For any diagonal ($n = m$) coefficient $C_{mn}$ one obtains the homogeneous differential equation

$$\partial^2 C_{mm}/\partial \varphi^2 + (m+\tau)^2 C_{mm} = 0\quad (8)$$

with homogeneous boundary conditions

$$\partial C_{mm}/\partial \varphi|_{\varphi=0} = 0,\quad \partial C_{mm}/\partial \varphi|_{\varphi=\pi} = 0.\quad (9)$$

Condition for the existence of a nonzero solution of (8), (9) when $m = 0$ leads to the characteristic equation for determination of the parameter $\tau$

$$\tau^2 \sin(\pi\tau) = 0.\quad (10)$$

The equation (10) has infinite set of nonnegative roots $\tau = 0, 1, 2, ...$ The dominant field behavior as $r \to 0$ corresponds to the lowest possible value $\tau = 0$. Note, that the root $\tau = 0$ is the triple one. Solving (8), (9) for $\tau = 0$, one can find

$$C_{mm} = p_{mm}\cos(m\varphi),\quad (11)$$

where $p_{mm}$ are constants.

Differential equation for $C_{mn}$, when $n < m$ becomes to be inhomogeneous. For example, if $n = m-1$ from (6) one can obtain the equation

$$\partial^2 C_{m,m-1}/\partial \varphi^2 + m^2 C_{m,m-1} = -2m^2 p_{mm}\cos(m\varphi),\quad (12)$$

solution of which is

$$C_{m,m-1} = p_{m,m-1}\cos(m\varphi)$$
$$+q_{m,m-1}\sin(m\varphi) - p_{mm}m\varphi\sin(m\varphi).\quad (13)$$

Here $p_{m,m-1}$ and $q_{m,m-1}$ are unknown constants. Substitution of (13) into (7), allows one to find

$$p_{mm} = -i\varsigma/(2\pi m^2 W_e) p_{m-1,m-1} \text{ or}$$
$$p_{mm} = -2iW_h/(\varsigma\pi m^2) p_{m-1,m-1} \quad (14)$$

for TM or TE polarization respectively. Here $1/W_e = 1/W_1 - 1/W_2$, $W_h = W_1 - W_2$. The recurrence relations (14) immediately determine the general expression for the diagonal coefficients

$$C_{mm} = (\varsigma/2\pi i W_e)^m p_{00} \cos(m\varphi)/(m!)^2 \text{ or}$$
$$C_{mm} = (2W_h/\pi i\varsigma)^m p_{00} \cos(m\varphi)/(m!)^2. \quad (15)$$

The formula (15) shows that the terms with $r \ln r$ and $r^2 \ln^2 r$, $r^3 \ln^3 r$, etc. have the same coefficient of proportionality $p_{00}$. It was claimed in [13, 14], that it is sufficient to restrict the logarithmic part in the expansion of function $\chi$ only by the term $r \ln r$. If one follows the suggestion of [13, 14] and excludes the term with $r^2 \ln^2 r$, then $p_{00} = 0$ should be accepted, and subsequently the whole solution vanishes. Therefore, the author's [13, 14] conclusion that the logarithmic term exists only linearly in the expression of the $\chi$ function is erroneous.

Further lowering of index $n$ leads to the sets of linear equations for sequential determination of unknown constants $p_{mn}$, $q_{mn}$. Since we are interested in the behavior of the fields as $r \to 0$, we shall write only some first terms in each component:

$$E_z = U_0 \left( 1 - \frac{i\varsigma \cos\varphi}{2\pi W_e} r \ln r - \frac{i\varsigma \sin\varphi}{2\pi} \right.$$
$$\left. \cdot \left( \frac{\pi - \phi}{W_1} + \frac{\varphi}{W_2} \right) r - \frac{\varsigma^2 \cos 2\varphi}{16\pi^2 W_e^2} r^2 \ln^2 r + ... \right),$$

$$H_\rho = U_0 \left( \frac{\sin\varphi}{2\pi W_e} \ln r + \frac{1}{2\pi} \left( \frac{\sin\varphi}{W_e} \right) \right.$$
$$\left. - \left( \frac{\pi - \varphi}{W_1} + \frac{\varphi}{W_2} \right) \cos\varphi \right) - \frac{i\varsigma \sin 2\varphi}{8\pi^2 W_e^2} r \ln^2 r + ... \right),$$

$$H_\varphi = U_0 \left( \frac{\cos\varphi}{2\pi W_e} \ln r + \frac{1}{2\pi} \left( \frac{\cos\varphi}{W_e} \right) \right.$$
$$\left. + \left( \frac{\pi - \varphi}{W_1} + \frac{\varphi}{W_2} \right) \sin\varphi \right) - \frac{i\varsigma \cos 2\varphi}{8\pi^2 W_e^2} r \ln^2 r + ... \right),$$

$$H_z = I_0 \left( 1 - \frac{2iW_h \cos\varphi}{\varsigma\pi} r \ln r - \frac{2i \sin\varphi}{\varsigma\pi} \right.$$
$$\left. \cdot ((\pi - \varphi)W_1 + \varphi W_2) r - \frac{W_h^2 \cos 2\varphi}{\varsigma^2 \pi^2} r^2 \ln^2 r + ... \right),$$

$$E_\rho = I_0 \left( -\frac{2W_h \sin\varphi}{\pi} \ln r + \frac{2}{\pi}((\pi - \varphi)W_1 + \varphi W_2) \cos\varphi \right.$$
$$\left. -\frac{2}{\pi} W_h \sin\varphi + \frac{2iW_h^2 \sin 2\varphi}{\varsigma\pi^2} r \ln^2 r + ... \right),$$

$$E_\phi = I_0 \left( -\frac{2W_h \cos\varphi}{\pi} \ln r - \frac{2}{\pi}((\pi - \varphi)W_1 + \varphi W_2) \sin\varphi \right.$$
$$\left. -\frac{2}{\pi} W_h \cos\varphi + \frac{2iW_h^2 \cos 2\varphi}{\varsigma\pi^2} r \ln^2 r + ... \right). \quad (16)$$

Here $U_0 = p_{00}$ for TM polarization and $I_0 = p_{00}$ for TE polarization. The obtained expressions describe following electromagnetic field's behavior near the common edge of two resistive half-planes:

$$E_z = O(1); \quad H_\rho, H_\varphi = O(\ln r);$$
$$H_z = O(1); \quad E_\rho, E_\varphi = O(\ln r). \quad (17)$$

It follows from (17) that for the junction of two resistive half-planes both transversal components $\mathbf{E}_\perp$ and $\mathbf{H}_\perp$ contain the logarithmic singularity simultaneously. The electromagnetic field induces a surface current on a resistive plane, and the density $\mathbf{j}$ of this current near the edge has the behavior $\mathbf{j} = O(1)$. Using (16) and according to the relations

$$E_x |_{y=0} = W j_x, \quad E_z |_{y=0} = W j_z \quad (18)$$

one can find that the $j_x$ component is continuous, while $j_z$ near the junction has a finite jump, which is proportional to $1/W_1 - 1/W_2$.

IV. CONCLUSIONS

The behavior of electromagnetic field near the common edge of two resistive half-planes with different surface impedances was investigated. We have rigorously proved that near the junction of half-planes both transversal components $\mathbf{E}_\perp$ and $\mathbf{H}_\perp$ contain the logarithmic singularity simultaneously. For the density of surface current, we have shown that the transversal component turned up to be continuous, while the longitudinal component near the junction has a finite jump, which is proportional to $1/W_1 - 1/W_2$. It is important to emphasize that involving to the expansion of the solution only the first power of logarithm in the term $r \ln r$ [13, 14] leads to the zero solution and so it is necessary to use also the higher power of logarithms in the terms $r^m \ln^n r$, $n \leq m$.